\documentclass[a4paper,fleqn,usenatbib]{mnras}

\usepackage[T1]{fontenc}
\usepackage{ae,aecompl}

\usepackage{graphicx}	
\usepackage{amsmath}	
\usepackage{amssymb}	
\usepackage{comment}
\usepackage{xcolor}
\usepackage{enumitem}
\usepackage{soul}
\usepackage{hyperref}

\newcommand{\ST}[1]{\textcolor{blue!100}{#1$_{\mathrm{ST}}$}}

\definecolor{steelblue}{rgb}{0.274 0.510 0.706}
\definecolor{mariogreen}{rgb}{0.165 0.616 0.561}

\title[Long-term dynamics of BBHs in YSCs]{Dynamics of binary black holes in 
young star clusters: the impact of cluster mass and long-term evolution} 

\author[S. Torniamenti et al.]{Stefano Torniamenti$^{1,2,3}$\thanks{E-mail: stefano.torniamenti@studenti.unipd.it}, Sara Rastello$^{1,2}$, Michela Mapelli$^{1,2,3}$\thanks{E-mail: michela.mapelli@unipd.it}, Ugo N. Di Carlo$^{4,3}$, 
\newauthor Alessandro Ballone$^{1,2,3}$, Mario Pasquato$^{1,2,5}$
\\
$^{1}$Physics and Astronomy Department Galileo Galilei, University of Padova, Vicolo dell'Osservatorio 3, I--35122, Padova, Italy\\
$^{2}$INFN - Padova, Via Marzolo 8, I--35131 Padova, Italy\\
$^{3}$INAF - Osservatorio Astronomico di Padova, Vicolo dell'Osservatorio 5, I-35122 Padova, Italy\\
$^{4}$McWilliams Center for Cosmology and Department of Physics, Carnegie Mellon University, Pittsburgh, PA, 15213,USA\\
$^{5}$Physics department, Montreal University, Montreal, Quebec H3T 1J4, Canada\\
}

\date{Accepted XXX. Received YYY; in original form ZZZ}

\pubyear{2021}

\begin{document}
\label{firstpage}
\pagerange{\pageref{firstpage}--\pageref{lastpage}}
\maketitle

\begin{abstract}
Dynamical interactions in dense star clusters  are considered one of the most effective formation channels of  binary black holes (BBHs). Here, we present direct $N-$body simulations of two different star cluster families: low-mass ($\sim{500-800}$ M$_\odot$) and relatively high-mass star clusters ($\ge{5000}$ M$_\odot$). We show that the formation channels of BBHs in low-  and high-mass star clusters are extremely different and lead to two completely distinct populations of BBH mergers. Low-mass clusters host mainly low-mass BBHs born from binary evolution, while BBHs in high-mass clusters are relatively massive (chirp mass up to $\sim{100}$ M$_\odot$) and driven by dynamical exchanges. Tidal disruption dramatically quenches the formation and dynamical evolution of BBHs in low-mass clusters on a very short timescale ($\lesssim{100}$ Myr), while BBHs in high-mass clusters undergo effective dynamical hardening until the end of our simulations (1.5 Gyr). In high-mass clusters we find that 8\% of BBHs have primary mass in the pair-instability mass gap,  all of them born via stellar collisions, while only one BBH with primary mass in the mass gap forms in low-mass clusters. These differences are crucial for the interpretation of the formation channels of gravitational-wave sources.

\end{abstract}




\section{Introduction}\label{intro}

Over the last six years, the LIGO \citep{aasi2015} and Virgo \citep{acernese2015} 
interferometers dectected an increasing number of gravitational wave (GW) events \citep[e.g.,][]{abbottO1,abbottGW150914,abbottO2,abbottGWTC2.1}. At the end of the third observing run, 
the third GW transient catalog (GWTC-3) consists of 90 GW candidates \citep{O3b,O3population}. 
Most of these events are produced by the inspiral of two 
black holes (BHs). Among the most peculiar  events, the merger remnant of GW190521 \citep{abbottGW190521,abbottGW190521astro} is the first  intermediate-mass BH (IMBH) ever detected in the mass range $100-1000 \, \mathrm{M_{\odot}}$, with a remnant mass of $142^{+28}_{-16} \, \mathrm{M_{\odot}}$. 
Also, GW190412 \citep[][]{abbottGW190412} represents the first observation of a BBH with asymmetric masses. The population of mergers in GWTC-3 also includes two binary neutron star (BNS) mergers, $\mathrm{GW170817}$ and  $\mathrm{GW190425}$ \citep{abbott170817mm,abbott170817grb,abbottGW190425}, and two black hole-neutron star (BHNS) candidates,  $\mathrm{GW200105\_162426}$ and $\mathrm{GW200115\_042309}$ \citep{abbottBHNS}. Furthermore, \cite{nitz2021} and \cite{olsen2022} reported several additional GW candidates \citep[see also][]{venumadhav2019,venumadhav2020,nitz2020,nitz2021}.

The abundance of detected GW sources allows us to attempt to reconstruct their formation channels. In fact, 
thanks to the distinctive features that different formation channels imprint on the merging progenitors, even a few hundreds detections may be sufficient to identify their main formation pathways \citep{fishbach2017, zevin2017,stevenson2017,farr2017,vitale2017,gerosa2017,gerosa2018,bouffanais2019,bouffanais2021a,bouffanais2021b,wong2019,wong2020,zevin2020,doctor2020,kimball2020,ng2020,roulet2021,mehta2022}. 
The isolated formation scenario, for example, predicts the formation of BBHs with primary masses up to $40-50 \, \mathrm{M_{\odot}}$, mostly equal-mass systems, with preferentially aligned spins and vanishingly small eccentricity in the LIGO--Virgo band \citep{mandel2016,gerosa2018}. According to this scenario, the formation of tight enough binary black holes (BBHs) can take place through evolutionary processes like common envelope \citep{tutukov1973, bethe1998,portegieszwart1998,belczynski2002, belczynski2008,belczynski2010,dominik2012, dominik2013,
mennekens2014, loeb16,belczynski2016,demink2016,marchant2016,
mapelli2018,mapelli2019, giacobbo2018b,
kruckow2018,spera2018,tang2019,belczynski2020,garcia2021}, chemically homogeneous evolution \citep{demink2016,mandel2016,marchant2016,dubuisson2020}, or stable mass transfer \citep[e.g.,][]{giacobbo2018,neijssel2019,bavera2020}. 

The dynamical formation scenario, instead, involves dynamical processes in dense stellar environments, 
like young star clusters \citep[YSCs, e.g.,][]{portegieszwart2002,sambaran10,mapelli2013,ziosi2014,goswami2014, mapelli2016, banerjee2017, banerjee2018,rastello2018,perna2019,
dicarlo2019,dicarlo2020,kumamoto2019,kumamoto2020,banerjee2020,rastello2020,rastello2021,dallamico2021,chattopadhyay2022}, 
globular clusters \citep[e.g.,][]{downing2010,BenacquistaDowning2013,rodriguez2015,rodriguez2016,antonini2016,
askar2017, fujii2017,askar2018,fragione2018,rodriguez2019,antonini2022}, and nuclear star clusters \citep[e.g.,][]{oleary2009,millerlauburg2009,mckernan2012,megan17,mckernan2018,vanlandingham2016,
stone2017,hoang2018,arcagualandris2018,antonini2018,arcasedda2019,arcasedda2020,mapelli2021}. 
With respect to the isolated channel, this scenario predicts the formation of merging BBHs with larger primary masses 
\citep[e.g.,][]{mckernan2012,mapelli2016,antonini2016,gerosa2017,stone2017,mckernan2018,dicarlo2019,dicarlo2020,rodriguez2019,yang2019,arcasedda2019,arcasedda2020,mapelli2021b}, 
isotropic spin distributions \citep[e.g.,][]{rodriguez2016b}, and, in some rare cases, non-zero eccentricity in the LIGO--Virgo band \citep[e.g.,][]{samsing2018,samsing2018b,samsing2018c,rodriguez2018,zevin2019,dallamico2021}.

In this work, we will focus on the dynamical formation of BBHs  in young 
and open stellar clusters. These systems are of key importance to interpret the formation of 
BBHs, because they host the formation 
of the most massive stars \citep{lada2003,portegieszwart2010,crowther2010}, which are the progenitors of massive BHs. Thanks to the high initial central density of the host cluster ($\rho \gtrsim 10^{3} \, \mathrm{M_{\odot} pc^{-3} }$, \citealp{portegieszwart2010}), binary stars can efficiently interact with the surrounding stars since the very beginning of their life. This leaves a deep imprint on the properties of the population of BBHs and, in turn, merging BBHs. When YSCs are eventually disrupted by the tidal field of their host galaxy, their stellar content is released into the galactic field. Thus, a large fraction of BBHs which are now in the field may have formed in young stellar systems.

The dynamical formation and evolution of BBHs in YSCs are explored in a realistic way by means of direct $N-$body simulations, where up-to-date prescriptions for single and binary stellar evolution are implemented. In many cases, only the first few hundreds Myr of the life of the star cluster are considered. This is due, on the one hand, to their small relaxation timescales, 
$t_{\mathrm{rlx}} \lesssim 100 \, \mathrm{Myr}$ \citep{portegieszwart2010}. In particular, the rapid decrease of their central density within the first Myr suppresses the interaction rate in later phases, making them less and less important in shaping the BBH properties. 
On the other hand, producing large sets of $N-$body simulations, which are necessary to explore the population of merging BBHs with sufficient statistics, has high computational costs. Thus, evolving large sets of YSCs for thousands of Myr would turn to be prohibitively expensive. As a consequence, many simulation of YSCs only take into account the first 100 Myr of the star cluster life \citep{,dicarlo2019,dicarlo2020,dicarlo2020b,rastello2020,rastello2021}.

The aim of this work is to evaluate the impact of the late phases (up to $1500 \, \mathrm{Myr}$) of the dynamical evolution of the star cluster on the population of BBH mergers. To this purpose, we have run two sets of $N-$body simulations of YSCs in different mass regimes and studied the evolution of the population of BBHs. 
To calculate the impact of the long-term evolution of the cluster, we compared the BBH merger populations at two different snapshots, $100 \, \mathrm{Myr}$ and $1500 \, \mathrm{Myr}$. The paper is organized as follows: in Section~\ref{sec_methods}, we describe the details of the $N-$body simulations and our method. In Section~\ref{sec_results}, we report the results for the BBH populations and mergers. Finally, Section~\ref{sec_conclusions} summarises our conclusions.

\begin{figure*}
\begin{center}
\includegraphics[width=\textwidth]{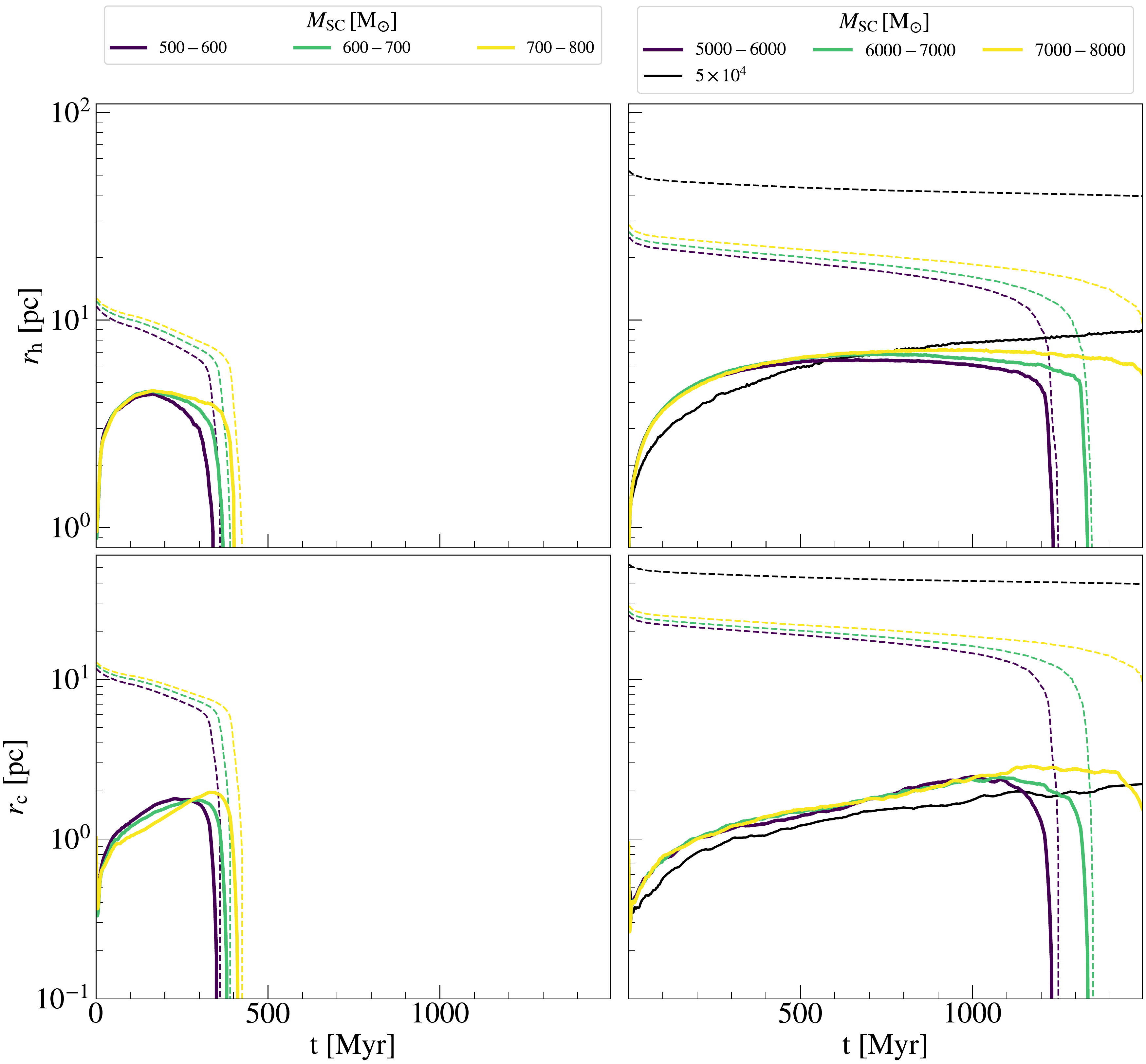}
\caption{
Evolution of the half-mass radius $r_{\mathrm{h}}$ (upper panels, solid lines), core radius $r_{\mathrm{c}}$ (lower panels, solid lines), and tidal radius $r_{\mathrm{t}}$ (dashed lines) for low-mass clusters (left) and high-mass clusters (right). Each set is divided into three subsets: for the low-mass clusters $M_{\mathrm{SC}} \in$ [500, 600] M$_\odot$ (violet), [600, 700] M$_\odot$ (green), [700, 800] M$_\odot$ (yellow). For the high-mass clusters $M_{\mathrm{SC}} \in$ [5000, 6000] M$_\odot$ (violet), [6000, 7000] M$_\odot$ (green), [7000, 8000] M$_\odot$ (yellow). Each line shows the median value over the simulated YSCs per each mass bin. The black lines (right panels) refer to the same physical quantities for the star clusters with $M_{\mathrm{SC}} =5\times10^4 \, \mathrm{M_{\odot}}$.
}\label{fig_radii}
\end{center}
\end{figure*}


\begin{figure}
\begin{center}
\includegraphics[width=\hsize]{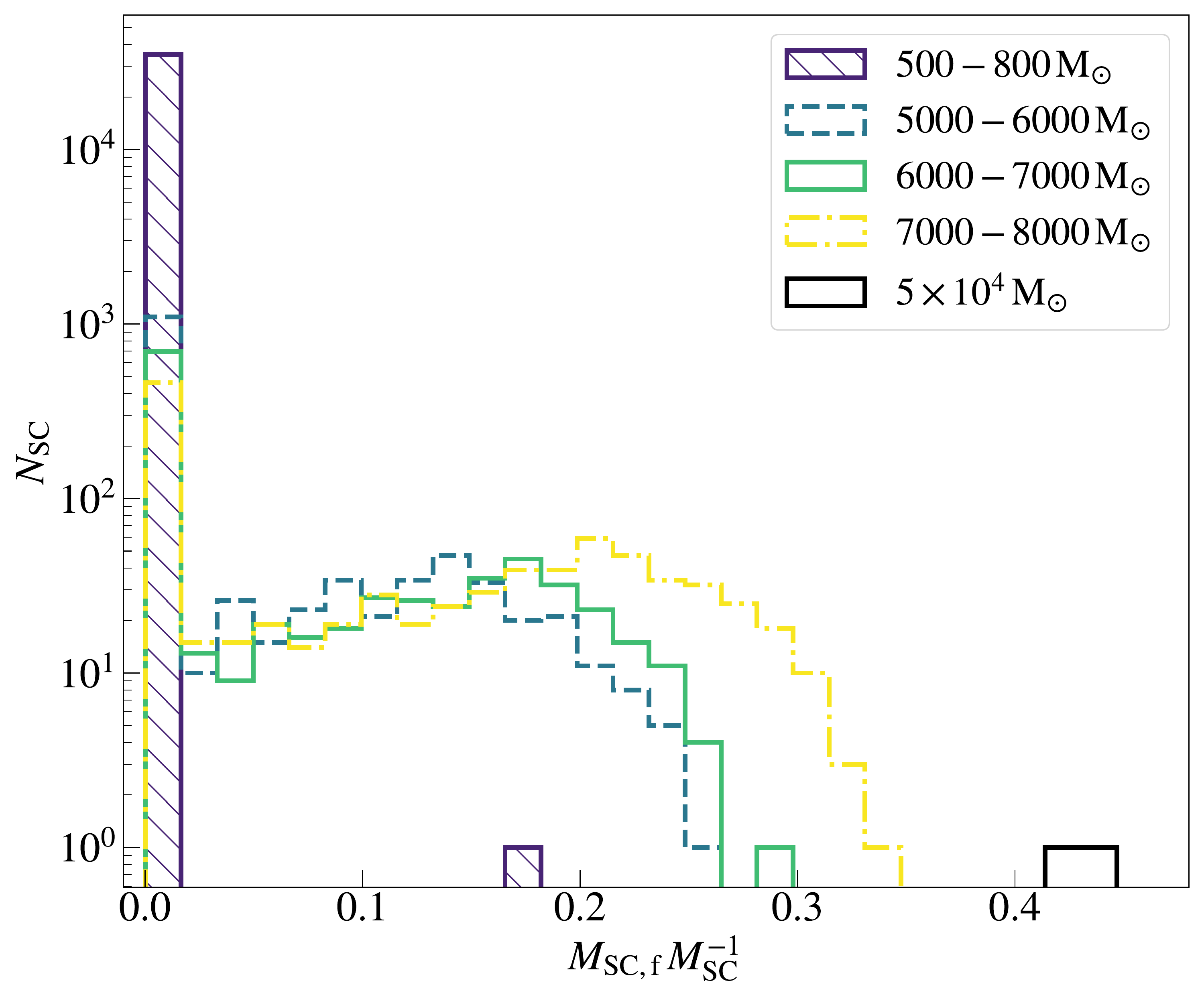}
\caption{Distribution of the ratio between the star cluster bound mass at the end of the simulation, $M_{\mathrm{SC,f}}$, and its initial mass $M_{\mathrm{SC}}$, for stellar systems with initial mass $M_{\mathrm{SC}} \in$ [500, 800] M$_\odot$ (violet), [5000, 6000] M$_\odot$ (blue), [6000, 7000] M$_\odot$ (green), [7000, 8000] M$_\odot$ (yellow), and  $M_{\mathrm{SC}} = 5 \times 10^4 \, \mathrm{M_{\odot}}$ (black).}\label{fig_mfin}
\end{center}
\end{figure}


\section{Methods}\label{sec_methods}


\subsection{Direct \textit{N}-body code} \label{sec_nbody}

We performed our simulations 
with the $N-$body code \textsc{nbody6++gpu} \citep{wang2015,wang2016}, coupled with the population synthesis code\footnote{{\sc mobse} is publicly available at \href{https://gitlab.com/micmap/mobse_open}{this link}.} \textsc{mobse} \citep{mapelli2017,giacobbo2018,giacobbo2018b}.
\textsc{nbody6++gpu} is the \text{GPU} parallel version of \textsc{nbody6} \citep{aarseth2003}. 
It implements a 4th-order Hermite integrator, individual block time-steps \citep{makino1992} and Kustaanheimo-Stiefel regularization of close encounters and few-body systems. The force contributions at short time steps (\textit{irregular forces}) are computed by a  neighbour scheme \citep{nitadori2012}, and for long time steps (\textit{regular} force/timesteps) the force is evaluated by considering all the particles of the system. The irregular-force calculation is performed using CPUs, while the regular forces are evaluated on GPUs using the CUDA architecture. 
A solar neighbourhood-like static external field \citep{wang2016} is included in the force integration. This choice for the tidal field is quite conservative, because the static tidal field does not take into account possible perturbations by disk and bulge shocking, and encounters with molecular clouds, which can accelerate the star cluster disruption \citep{gieles2006}.
Orbital decay and circularization by GW emission are calculated following \cite{peters1964}.
 Post-Newtonian terms are not 
included in this version of {\sc nbody6++gpu}.

\textsc{mobse} is a customized and upgraded version of \textsc{bse} and includes up-to-date prescription for massive stellar winds \citep{giacobbo2018}, core-collapse \citep{fryer2012} and electron-capture supernovae \citep{giacobbo2018c}, natal kicks \citep{giacobbo2020} and (pulsational) pair instability supernovae \citep{mapelli2020b}. 
Stellar winds are modeled by assuming that the mass loss of hot massive stars depends on metallicity as $\Dot{M} \propto Z^{\beta}$, where $\beta$ is modelled as in \cite{giacobbo2018}.

For this work, we adopt the delayed model for core-collapse supernovae from \cite{fryer2012}. In this model, there is no mass gap between neutron stars and BHs: we assume that compact objects more massive than 3 M$_\odot$ are BHs. Natal kicks are modeled according to the prescription by  \cite{giacobbo2020}: the magnitude of the kick can be expressed as $v_{\mathrm{kick}} \propto f_{\mathrm{H05}} \, m_{\mathrm{ej}} \, m_{\mathrm{rem}}^{-1}$, where $f_{\mathrm{H05}}$ is a random number drawn from a Maxwellian distribution with a one-dimensional root mean square velocity $\sigma = 265 \, \mathrm{km \, s^{-1}}$,  $m_{\mathrm{rem}}$ is the mass of the remnant, and $m_{\mathrm{ej}}$ is the difference between the final mass of the star before the supernova explosion and the mass of the remnant. Binary evolution processes (tides, mass transfer, common envelope and GW-orbital decay) are implemented as in \cite{hurley2002}. The common envelope process is implemented by adopting the energy formalism \citep{webbink1984}. In this case, we assume $\alpha=3$, while the concentration parameter $\lambda$ is calculated self-consistently as in \cite{claeys2014}.

\subsection{Initial conditions} \label{sec_initial_coditions}


\begin{figure*}
\begin{center}
\includegraphics[width=\textwidth]{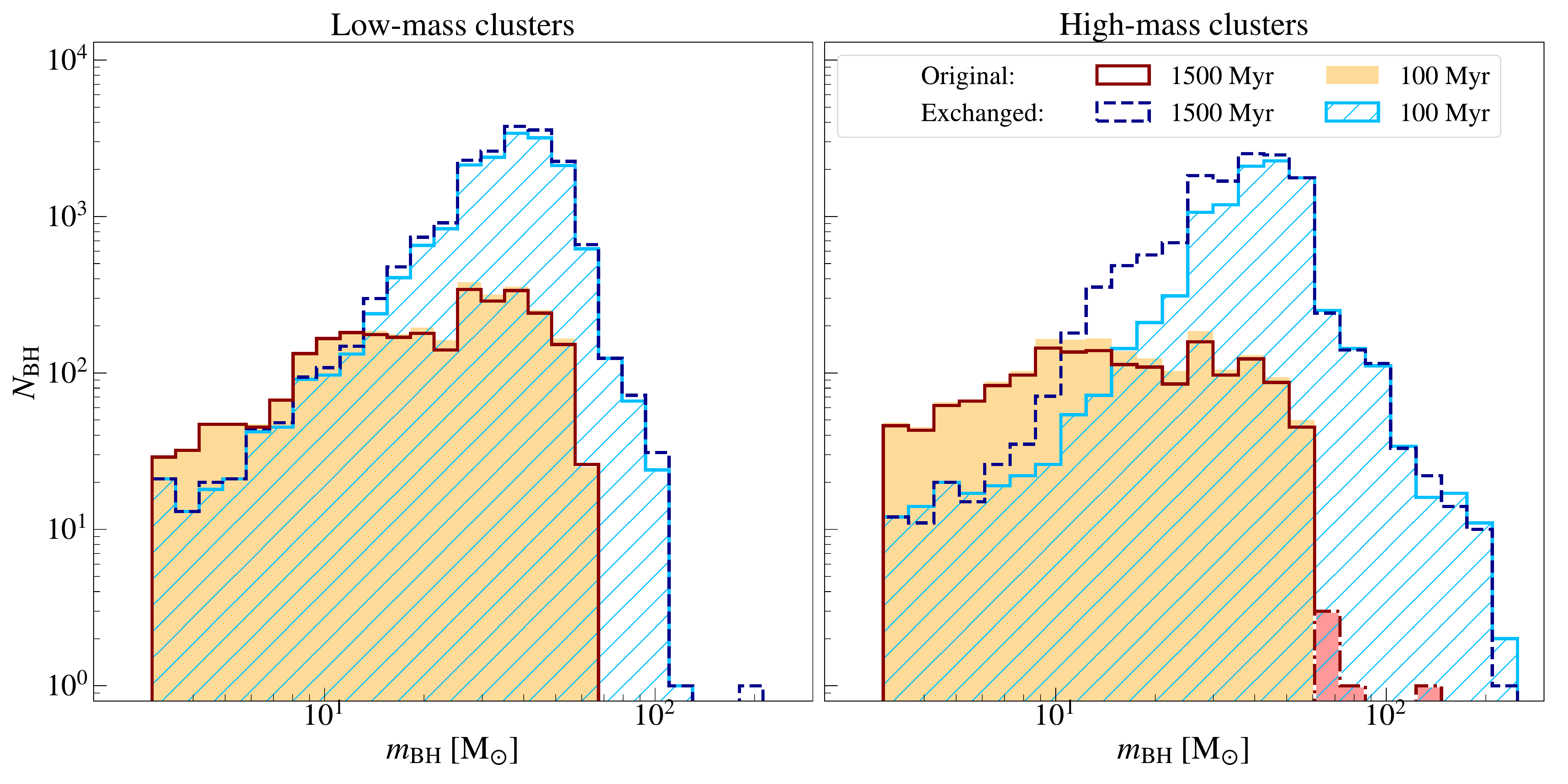}
\caption{Mass distribution of BHs in BBHs, for low-mass clusters (left) and high-mass clusters (right). 
Red line: original BBHs at 1500 Myr.  Orange filled histogram: original BBHs at 100 Myr. Blue dashed line: exchanged BBHs at 1500 Myr.  Light blue hatched histogram: exchanged BBHs at 100~Myr.
Among the original BHs, we highlight in red those with mass in the PI mass gap at 1500 Myr (red dash-dotted line) and at 100 Myr (red filled histogram). These are anomalous original BBHs in which one of the two stellar components has merged with another star before producing the BBH (see Section~\ref{sec_bbhl} for details).}\label{fig_all_bbh}
\end{center}
\end{figure*}


\subsubsection{Stellar and binary populations}

We generate the initial masses of stars (single stars, primary and secondary members of binary systems)  according to a \cite{kroupa2001} initial mass function between $0.1 \, \mathrm{M_{\odot}}$ and $150 \, \mathrm{M_{\odot}}$. We assume a metallicity $Z=0.002$, approximately corresponding to $0.1$ Z$_\odot$. 
For binary systems, we assume a distribution of mass ratios  $\mathcal{F}(q) \propto q^{-0.1}, \;  \mathrm{with} \; q \; \in [0.1,\,{}1]$ \citep{sana2012}.

{ Our algorithm generates a mass-dependent binary fraction 
 $f_{\rm b}$, according to \cite{moe2017}\footnote{ \cite{moe2017} take their data from different spectroscopic surveys, including binary systems in the field, OB associations, and young star clusters. Ideally, here we should consider only the sub-sample of binary systems in young star clusters \citep{sana2012}, but this is not feasible, because the lowest-mass stars are missing in such sub-sample.}. 
The details of the assumed binary fraction per mass bin are shown in Table~\ref{tab:binfrac}.  }
We generate the orbital parameters of binary systems following the observational prescriptions by \cite{sana2012}. 
In particular,  we randomly draw the orbital periods from: $\mathcal{F}(\mathcal{P}) \propto \mathcal{P}^{-0.55}, \; \mathrm{with} \; \mathcal{P}=\log_{10}(P/ {\rm days})\in [0.15,\,{}5.5], $
and the eccentricities from $\mathcal{F}(e) \propto e^{-0.45}, \; \mathrm{with} \; e \in [10^{-5},\,{}e_{\rm max} (P)]
$. 
For a given orbital period, we set an upper limit for the eccentricity distribution according to \cite{moe2017}: $e_{\rm max} (P) = 1 - \left[ P / (2 \, {\rm days}) \right]^{-2/3}$. { Our method allows to obtain orbital properties for O-type stars (i.e., the progenitors of BHs) consistent with those observed in young and open clusters and OB associations (\citealp{sana2011,sana2012}). Based on population-synthesis simulations by \cite{deminkbelczynski2015}, we expect that our choice of the initial binary parameters has a mild (negligible) impact on the evolution of BBHs in low-mass (high-mass) clusters, where binary evolution is most (least) important with respect to dynamical interactions. 
We refer to \cite{torniamenti21} for more details on our initial binary population.}

\begin{table}
  \centering
    \begin{tabular}{cc}
    \hline
           Mass Bin [M$_\odot$] & Binary fraction $f_{\rm b}$  \\
          \hline\hline
    0.1--0.8 & 0.2\\
    0.8--2.0 & 0.4\\
    2.0--5.0 & 0.59\\
    5.0--9.0 & 0.76\\
    9.0--16.0 & 0.84\\
    16.0--150.0 & 0.94\\
    \hline
    \end{tabular}  
  \caption{ Adopted values of the original binary fraction $f_{\rm b}$ (column 2) per each stellar mass bin (column 1). From \protect\cite{moe2017}.}\label{tab:binfrac}
\end{table}%



\subsubsection{Stellar clusters} \label{sec_star_clusters}

We initialize stellar positions and velocities in the simulated  YSCs 
with fractal initial conditions, with a fractal dimension $D=1.6$, in order to mimic the observed clumpiness of embedded star clusters \citep{cartwright2004,sanchez2009, kuhn2019}.  We generate fractal phase space distributions with {\sc Mcluster} \citep{kuepper2011}.

We uniformly sample the half-mass of our star clusters  between 0.5 and 2 pc  \citep[e.g.,][]{portegieszwart2010,krumholz2019}. 
To evaluate the impact of long-term dynamics on the properties of BBH mergers in different dynamical regimes, we consider two sets of star clusters in different mass ranges ($M_{\mathrm{SC}}$):
\begin{itemize}
    \item {\emph{Low-mass star clusters}}, with mass ranging from $500 \, \mathrm{M_{\odot}}$ to  $800 \, \mathrm{M_{\odot}}$. These clusters present short dynamical evolution timescales at all scales: this reduces the probability of dynamical interactions and, consequently, of dynamical exchanges. Also, YSCs in this mass range typically host a few massive stars, and, consequently, BHs. This further suppresses the rate of dynamical exchanges\footnote{  
    An exchange tends to happen when the intruder is more massive than at least one of the members of the binary system, because this leads to a substantial increase of the binary's binding energy, hardening the binary system  \citep{heggie1975,hills1980,heggie2003}. Our low-mass clusters lack massive intruders because all of the massive stars are already born in hard binary systems. Thus, exchanges tend to be suppressed.} \citep{rastello2021}.   
    
    \item {\emph{High-mass star clusters}}, with mass ranging from $5000 \, \mathrm{M_{\odot}}$ to $8000 \, \mathrm{M_{\odot}}$. These clusters have a higher rate of dynamical encounters, as a consequence of the higher densities, 
    longer dynamical timescales and 
    larger number of massive stars. Thus, they are expected to produce a larger number of exchanged binaries and BBHs (\citealp{rastello2021}).  In this sample we also include two star clusters with mass $5\times 10^4 \, \mathrm{M_{\odot}}$. 
 
\end{itemize}
{ Here, the terms \emph{low-mass} and \emph{high-mass} clusters are intended only for comparison between the two considered samples. In the literature, low-mass clusters can include even star clusters with much lower mass, down to a few ten M$_\odot$ \citep[e.g.,][]{lada2003}, while the highest mass clusters can reach $\sim{10^7}$ M$_\odot$ \citep[e.g.,][]{georgiev2016}.}

In both cases, we sample the mass of star clusters from a power-law distribution $dN/dM_{\mathrm{SC}} \propto M_{\mathrm{SC}}^{-2}$, following \cite{lada2003}. The two sets consist in 35578 and 3555 star clusters, respectively. The number of star clusters in the two samples is set to obtain the same total mass. The total kinetic ($K$) and potential ($W$) energy of the cluster are set to give a virial ratio $Q=2\,{}K/W=1$.

\begin{figure*}
\begin{center}
\includegraphics[width=\textwidth]{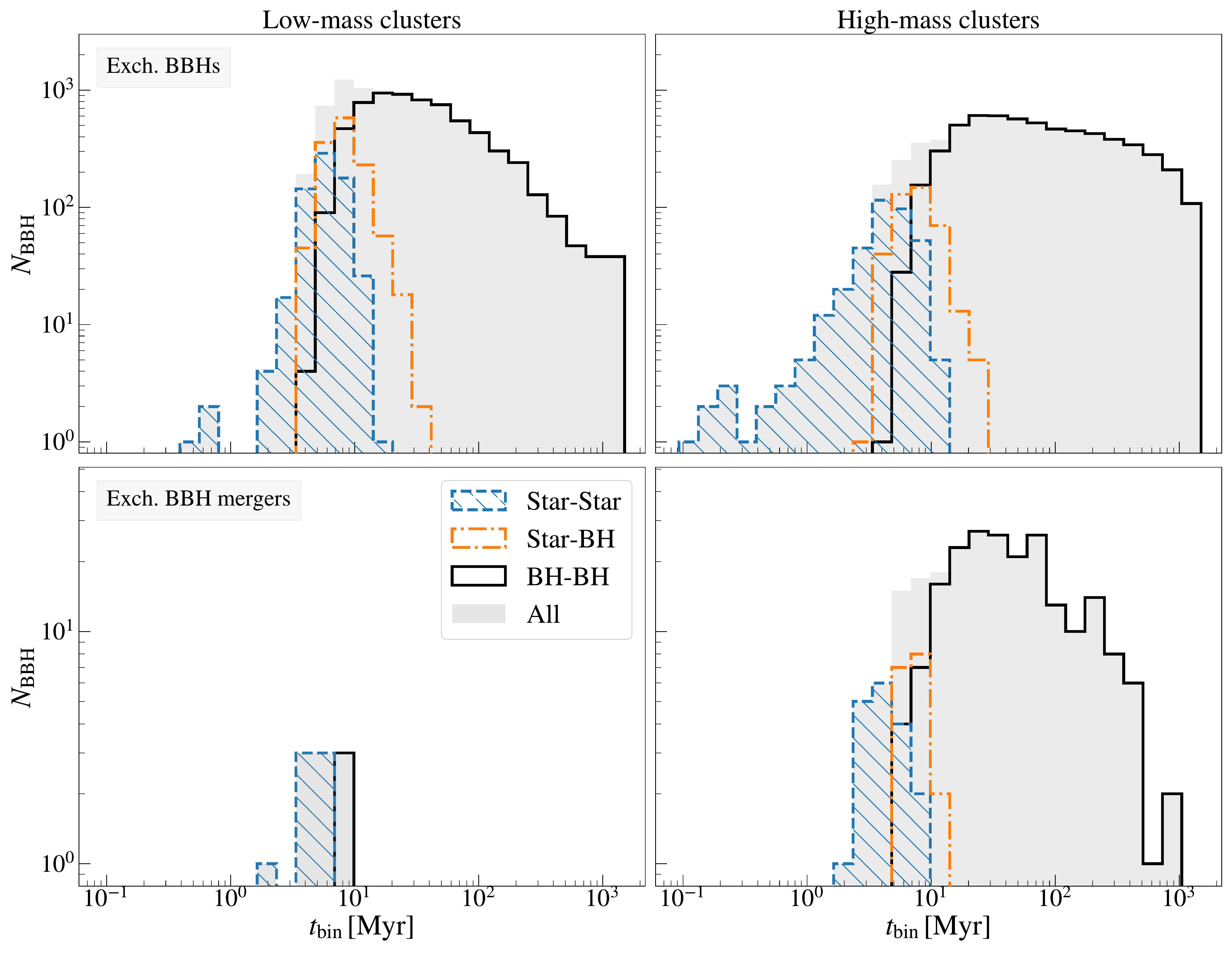}
\caption{{ Distribution of formation times ($t_{\mathrm{form}}$) of binary systems that give birth to exchanged BBHs, in low-mass clusters (left) and high-mass clusters (right). Upper panels: all exchanged BBHs. Lower panels: exchanged BBH mergers. Blue dashed line and hatched area: BBHs that formed when both components were stars. Orange dot-dashed line: BBHs that formed when one component was a star and the other was a BH. Black line: BBHs that formed when both components were BHs. Grey area: all BBHs.}}\label{fig_bin_form}
\end{center}
\end{figure*}


\subsection{Impact of long-term evolution}

The main goal of this work is to evaluate the impact of 
{  dynamics on the population of BBH mergers. In particular, we simulate our star clusters} 
up to $t_{\rm f}=1500 \, \mathrm{Myr}$.
{ This corresponds to a total integration time of $\sim 150$ relaxation timescales ($t_{\mathrm{rlx}}$,  \citealp{spitzer1987}) for low-mass clusters which have, on average,  $t_{\mathrm{rlx}} \sim 10$ Myr. 
For high-mass cluster, on average, $t_{\mathrm{rlx}} \sim 26$ Myr, and the total integration time is longer than $ 50 \, t_{\mathrm{rlx}}$.} { To be completely sure that we captured  all the dynamical encounters relevant for the population of BBHs, we should integrate all our clusters until they become tidally filling, or until the last BH leaves its parent cluster by ejection or evaporation. For the most massive clusters ($>7000$ M$_\odot$), our clusters become tidally filling at later times with respect to $t_{\rm f}=1.5$ Gyr (Figure~\ref{fig_radii}). However, integrating our most massive clusters for more than 1.5 Gyr currently requires prohibitive computational resources. Moreover, we found that only 0.01\% (6\%) of all BBHs are still bound to their parent cluster in our low-mass (high-mass) star clusters. Hence, the vast majority of our BBHs has already been dynamically ejected at 1.5 Gyr and will not be affected by dynamical interactions at later stages.} 

We compare the population of BBH mergers that form in the first 100 Myr of the evolution of the simulated YSCs with the population of BBH mergers at 1500 Myr.
In particular, we first evaluate the population of BBH mergers that we would have obtained if we had integrated the evolution of our YSCs only for the first 100 Myr. 
This population consists in: 
\begin{itemize}
    \item BBHs that merge within the first 100 Myr, during the $N-$body simulations.
    \item BBHs that will merge within a Hubble time \textbf{in absence of further dynamical interactions}. To calculate them, we consider the population of existing BBHs at 100 Myr and evolve their orbital eccentricity and semi-major axis by integrating the equations of \cite{peters1964}, to calculate the energy loss due to GW emission:
    \begin{eqnarray}
       \frac{{\rm d}a}{{\rm d}t}=-\frac{64}{5}\,{} \frac{G^3 \,{} m_1 \,{} m_2 \,{} (m_1+m_2)}{c^5 \,{} a^3\,{} (1-e^2)^{7/2}}\,{}f_1(e), \nonumber\\
  \frac{{\rm d}e}{{\rm d}t}=-\frac{304}{15}\,{} e \frac{ G^3 \,{} m_1 \,{} m_2 \,{} (m_1+m_2)}{c^5 \,{}a^4 \,{}  (1-e^2)^{5/2}}\,{}f_2(e),\nonumber\\
    \end{eqnarray}
    where
    \begin{eqnarray}
     f_1(e)=\left(1+\frac{73}{24}\,{}e^2+\frac{37}{96}\,{} e^4\right) \nonumber\\
f_2(e)=\left(1+\frac{121}{304} \,{} e^2\right).
    \end{eqnarray}
    
    All the BBHs that merge within a Hubble time ($t_{\mathrm{H}}= 14 \, \mathrm{Gyr}$) are classified as mergers. This is equivalent to assume that the YSCs dissolve at 100 Myr, and their BBHs evolve only via GW emission (no dynamical interactions) after the death of their parent star clusters.
\end{itemize}

To evaluate how dynamical encounters affect the distribution of BBH mergers in the late phases of the cluster life, we repeat the aforementioned procedure after 1500 Myr, i.e. \begin{itemize}
    \item we count how many BBHs merge within 1500 Myr, during the $N-$body simulations;  
    \item we integrate the semi-major axis and eccentricity evolution of the other BBHs that are still bound at 1500 Myr, accounting for GW emission only \citep{peters1964}, and we count how many of them merge within one Hubble time. 
\end{itemize}

If the dynamical interactions within the cluster are still effective after 100 Myr, they can affect the population of BBHs, and, consequently, of BBH mergers. 
In particular, dynamical processes can form new BBHs or harden the existing ones, allowing them to merge within an Hubble time, or even before the end of the simulation, thus increasing the population of BBH mergers. In some other cases, dynamical interactions can disrupt existing BBHs, possibly removing them from the population of mergers that we estimated at 100 Myr.

\subsection{Estimate of relativistic kicks} \label{sec_rel_kick}
{  When two BHs merge, the post-merger remnant receives a kick 
due to the asymmetric momentum dissipation by  GWs \citep[e.g.,][]{favata2004}. This recoil can 
reach up to thousands km s$^{-1}$, depending on the symmetric mass ratio and spin orientation (\citealp{campanelli2007}).
If the kick magnitude is larger than the escape velocity of the host star cluster, the post-merger remnant is ejected.
For the star clusters considered in this work (Sect. \ref{sec_initial_coditions}), the initial escape velocities range from $\sim{1}$ to $\sim{10}$ km s$^{-1}$, and may rapidly decrease to zero as a consequence of cluster dissolution. 

As our simulations do not include relativistic kicks, we evaluated the probability that a post-merger remnant is ejected  a posteriori, using the equations reported by  \cite{maggiorebook}. 
In particular, for each BBH merger, we randomly draw a distribution of spin magnitudes for each component from a Maxwellian distribution: we consider two cases: $\sigma_\chi{} =0.1$ and $\sigma_\chi{} =0.01$ \citep[see, e.g.,][for this assumption]{bouffanais2021b}.  
Also, we assume that the spin directions are isotropically distributed over the sphere \citep[e.g.,][]{rodriguez2016b}. We randomly draw $10^5$ different spin magnitudes and orientations for each BH, and calculate the resulting relativistic kick distribution. Then, we evaluate the probability that the BH remnant is retained within the cluster, that is the probability to find a kick value lower than the escape velocity at the BBH merger.  

For all our BBH mergers, we find a retention probability $p<2\%$ if $\sigma_\chi{} =0.1$ and $p<5\%$ if $\sigma_\chi{} =0.01$. Thus, we can safely assume that all the post-merger remnants are ejected from their host cluster. Therefore, we will remove from our sample 
any second-generation BBH, 
i.e. any BBH that has at least one component resulting from a previous BBH merger.
}

\begin{figure*}
\begin{center}
\includegraphics[width=\textwidth]{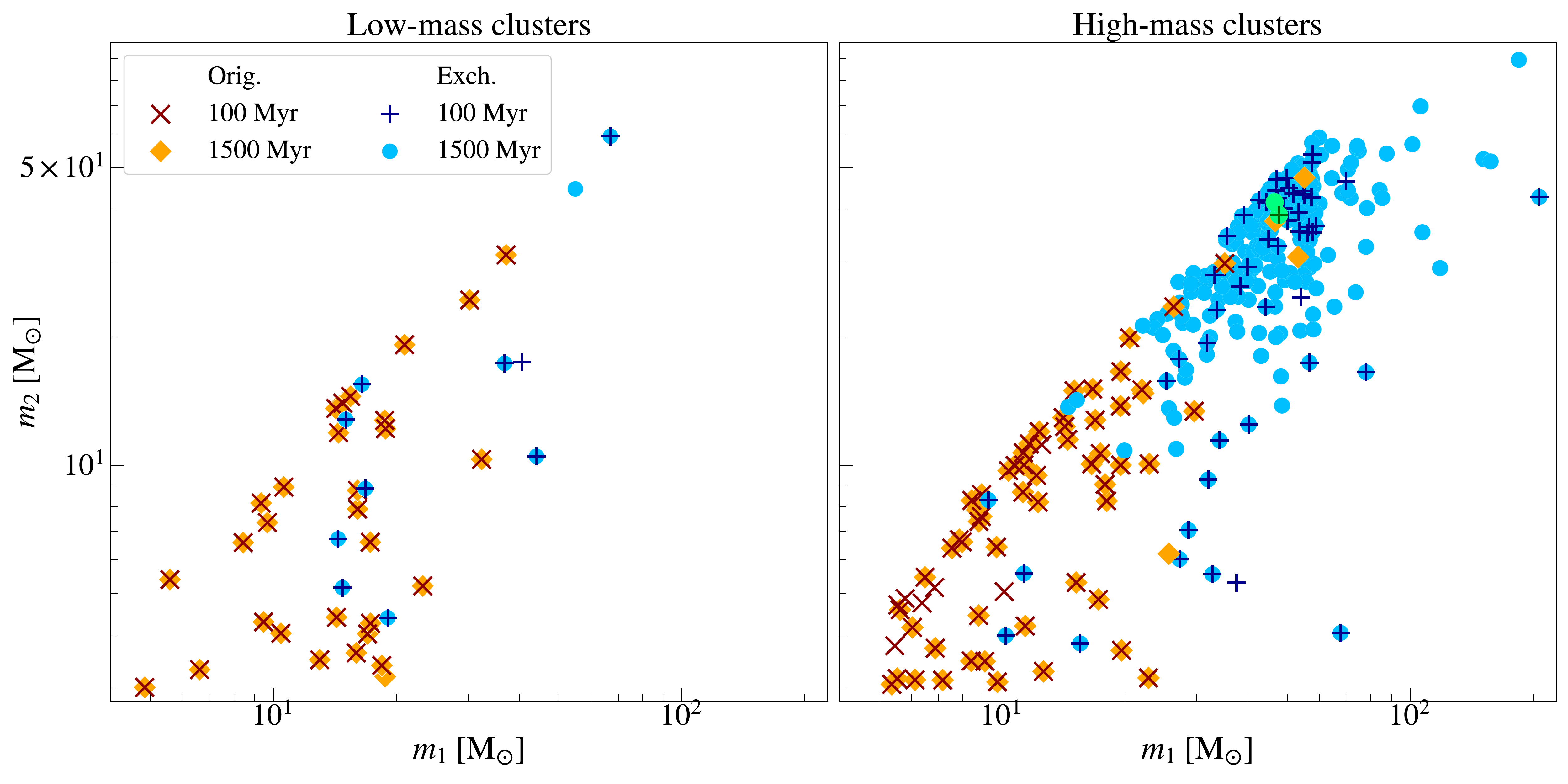}
\caption{Mass of the secondary BH ($m_{\mathrm{2}}$) versus the primary BH  ($m_{\mathrm{1}}$) of  { BBH mergers} in low-mass clusters (left) and high-mass clusters (right). Orange diamonds: original BBHs at 1500 Myr. Red crosses: original BBHs at 100 Myr. Light blue circles: exchanged BBHs at 1500 Myr. Blue plusses: exchanged BBHs at 100 Myr. Green circles: exchanged BBHs at 1500 Myr in star clusters with mass $5 \times 10^{4} \, \mathrm{M_{\odot}}$.}\label{fig_bbh_merger}
\end{center}
\end{figure*}


\section{Results}\label{sec_results}

\subsection{Global evolution of the cluster} \label{sec_global}

Figure \ref{fig_radii} shows the evolution of the half-mass radius $r_{\mathrm{h}}$, tidal radius $r_{\mathrm{t}}$, and core radius $r_{\mathrm{c}}$ of the two sets of clusters. 
Each set is split into three sub-sets of different mass in order to better take into account the impact of the cluster mass on its expansion. As a comparison for high-mass clusters, we also show the evolution of the stellar clusters with $5 \times 10^4 \, \mathrm{M_{\odot}}$.

{ All the clusters show an initial rapid expansion due to the early mass loss caused by stellar and binary evolution  (stellar winds and supernovae).
This expansion is visible at all scales, but is more pronounced for $r_{\mathrm{h}}$, consistently with  the results of \cite{chattopadhyay2022}. Also, the early mass loss causes the initial steep decrease of the tidal radius. 

After the most massive stars have evolved into compact remnants, the cluster enters a slower, relaxation-driven phase, as visible from the change of slope of $r_{\mathrm{t}}$. 
In particular, the tendency towards energy equipartition makes the low-mass stars move altogether to the outer regions, approaching the tidal boundary, while the heavier stars and compact remnants sink at the centre of the cluster, if they are not already segregated (\citealp{spitzer1987}). As a result, stars are progressively removed by the galactic tidal field, 
leading to cluster dissolution. Eventually, the tidal stripping, further enhanced by the cluster mass loss, and the possible energy generation by massive BBHs within the cluster core lead the stellar cluster out of dynamical equilibrium (\citealp{giersz2019}), causing an abrupt disruption of the cluster and a sharp decrease of its relevant radii.

The duration of the relaxation-driven phase depends on the initial cluster mass. For low mass clusters, the short relaxation timescales (\citealp{spitzer1987}), combined with the small initial mass, cause a rapid dissolution. Their typical lifetime is 400 Myr, with no particular distinction among the three subsets. 
The evolution of high-mass clusters, instead, is characterized by a much milder expansion, where the growth of $r_{\mathrm{h}}$ is balanced by  tidal stripping. As a result, $r_{\mathrm{h}}$ remains almost constant (a similar trend can be seen in \citealp{banerjee2017} for more massive clusters) for the first 1000 Myr. 
The cluster dissolution happens at different times, depending on the mass subset considered. The typical lifetime  spans from $1250 \, \mathrm{Myr}$ for clusters with mass $<6000 \, \mathrm{M_{\odot}}$, up to more than $1500 \, \mathrm{Myr}$ for clusters more massive than $7000 \, \mathrm{M_{\odot}}$. In contrast, the $5 \times 10^4 \, \mathrm{M_{\odot}}$ clusters undergo a far slower expansion and their tidal radius is almost unchanged at the end of the simulation, except for the initial rapid decrease due to stellar evolution mass loss. 

In both low- and high-mass clusters, the core radius increases monotonically until the cluster is disrupted, thus lacking a clear core-collapse phase. As shown by \cite{chattopadhyay2022}, this is not unexpected for models with a large primordial binary population. In this case, primordial binaries can heat up the cluster since the very beginning of its evolution, preventing a deep core collapse.
}


Figure \ref{fig_mfin} shows the distribution of the ratio between the star cluster bound mass at 1500 Myr, $M_{\mathrm{SC,f}}$, and its initial mass, $M_{\mathrm{SC}}$. Low-mass clusters are completely disrupted by the tidal field of the host galaxy at the end of the simulation. { In only one case, a core of about $120 \, \mathrm{M_{\odot}}$, corresponding to 18\% of the initial mass, can survive\footnote{ Most of the remaining mass of this interesting survivor consists of a BBH with a total mass of 80 $\mathrm{M_{\odot}}$, composed of two 
BHs of 44 $\mathrm{M_{\odot}}$ and 36 $\mathrm{M_{\odot}}$, respectively. This BBH formed via dynamical exchange. The rest of the mass is distributed in $50$ low-mass stars. This is a unique case in our sample. We may speculate that, because the BBH formed in a relatively loose environment ($\sim 10 \, \mathrm{M_{\odot} \, pc^{-3}}$), its high mass and the poor dynamical interaction rate allowed a number of stars to remain within the tidal radius, without being scattered away by the BBH itself. 
}.}  

As for high-mass clusters, one third of the stellar systems are still bound at the end of the simulation. In this set, the number of surviving clusters increases with the initial mass of the cluster. Also, more massive clusters can generally retain a higher fraction of mass. Finally, stellar clusters with $5 \times 10^4 \, \mathrm{M_{\odot}}$ preserve about half of their initial mass at 1500 Myr.

\subsection{BBH populations} \label{sec_bbhl}

Figure \ref{fig_all_bbh} shows the mass distribution of BHs in BBHs in the two considered snapshots: 100 Myr and 1500 Myr.  { In this section, we analyse
both merging and non-merging BBHs.} Among the BBH populations, we distinguish between \textit{original BBHs}, whose progenitors were already present as binary stars in the initial conditions of the simulation\footnote{ In the literature, a binary star that is already present in the initial conditions of a direct $N$-body simulation is often referred to as a primordial binary star. Here, we use the term \emph{original} instead of \emph{primordial}, to avoid any possible confusion with the concept of primordial BHs \citep[e.g.,][]{carr2016}.}, and \textit{exchanged BBHs}, which have formed as a consequence of dynamical exchanges.

In low-mass clusters, the BH populations in the two snapshots are almost identical, suggesting that, after $100 \, \mathrm{Myr}$, dynamical encounters play a negligible role in the evolution of BBHs. 
In contrast, high-mass clusters are still { dynamically} active 
at later phases, as indicated by the 
increase of { the number of } exchanged BHs between  $7 \, \mathrm{M_{\odot}}$ and $50 \, \mathrm{M_{\odot}} $. 

{ The BH populations of low- versus high-mass clusters show several differences.} 
First, 
low-mass clusters display a larger number of original BBHs with $m_{\mathrm{BH}}> 20 \, \mathrm{M_{\odot}}$. 
{  Such massive original BBHs tend to be suppressed in high-mass clusters, because 
they undergo stronger dynamical interactions for a longer time, eventually resulting in a dynamical exchange. 
This leads to 
the formation of nearly equal-mass, massive exchanged BBHs in the high-mass clusters. This process is also enhanced by the higher initial number of massive stars in high-mass clusters 
with respect to low-mass clusters. As a consequence, original BBHs in high-mass clusters tend to have lower masses than those in low-mass clusters. }

Figure \ref{fig_bin_form} displays the distribution of formation times of the binary systems that give birth to exchanged BBHs ($t_{\mathrm{form}}$). In low-mass clusters, 8\% of these systems form when  both components are still stars, and about 15\% when only one component is a BH. 
Most of the { binaries that result in BBHs} form by the pairing of two BHs, with a peak { at formation time} $t_{\rm form}=10-20$ Myr. The rapid decrease of the dynamical activity and the dissolution of the stellar cluster cause a steep decrease in the distribution of $t_{\mathrm{form}}$, and only 12\% of the BBHs form after 100 Myr { in low-mass clusters}. 

In high-mass clusters, about 88\% of the { binaries resulting in BBHs} form from the pairing of two BHs. In this case, the distribution of $t_{\mathrm{form}}$ shows a flatter trend, hinting at an efficient dynamical activity of the cluster at later times. { In fact,} more than one third of the BBHs { pair up} after 100 Myr. In these clusters, only 5\% of the binaries that produce BBH systems form when  both components are still stars. 

\subsubsection{Mass-gap BHs and IMBHs}

{ 
More than 7\% of our simulated BBHs in high-mass clusters (here we consider both merging and non-merging systems) have primary mass $>60 \, \mathrm{M_{\odot}}$. 
}

{ The formation of such massive BHs, with mass ranging from $\sim 60 \, \mathrm{M_{\odot}}$ to $\sim 120 \, \mathrm{M_{\odot}}$ is suppressed in single stellar { evolution} by pair-instability (PI) and pulsational pair instability (PPI). Nonetheless, as shown by \cite{spera2018}, \cite{dicarlo2019} and \cite{dicarlo2020}, BHs in the mass gap can form as a consequence of stellar mergers, which produce very massive stars that eventually collapse to BHs. }

{ BHs in this mass range may also be the result of previous BH mergers (e.g., \citealp{banerjee2021}). However,  almost all merger remnants are expected to be ejected by  relativistic kicks in our simulated star clusters (Sect. \ref{sec_rel_kick}). 
Hence,  all the BHs which have a mass $60-120$ M$_\odot$  and are members of BBHs form as a result of stellar mergers 
in our simulated star clusters.} 

{ In five cases, a BH with mass $>60$ M$_\odot$ even forms in an original binary system. These are systems in which the original binary remains bound after the merger of one of its components with a third star, producing an original binary with a mass-gap primary BH (Fig.~\ref{fig_all_bbh}).}

{ Low-mass clusters display a lower percentage of mass-gap BHs in BBHs ($\sim 4 \%$), because of their lower rate of dynamical interactions as well as the limited initial number of massive stars (Sect. \ref{sec_bbhl}). Also, no mass-gap BH is present in original binary systems.
} 

{ Furthermore,  $\sim{1.5}$\% of all BHs that are bainry members in our high-mass clusters are IMBHs (i.e., BHs with mass $>100$ M$_\odot$). The maximum BH mass we find in BBHs in our high-mass clusters is 250 M$_\odot$. As in the case of BHs in the PI mass-gap, all the IMBHs we found form via multiple stellar collisions \citep[e.g.,][]{dicarlo2021}.}
 { In contrast, 
only eight BHs with mass $>100 \, \mathrm{M_{\odot}} $ form in low-mass clusters, { corresponding to $\sim{0.1}$\%  of all the BHs born in the low-mass clusters}. }

\subsection{BBH mergers} \label{sec_mergers}

{ In this Section, we focus on BBH mergers, i.e. BBHs that reach coalescence in less than 14 Gyr.}

\subsubsection{Low-mass clusters}

Figure \ref{fig_bbh_merger} shows the mass of the secondary BH ($m_{\mathrm{2}}$) versus the primary BH ($m_{\mathrm{1}}$) for BBH mergers. In low-mass clusters, the population of BBH mergers mostly consists of original BBHs, as a further proof of the poor dynamical activity of these systems. In general, dynamical exchanges do not affect the population of BBHs after 100 Myr { ($\sim 10 \, t_{\mathrm{rlx}}$)}, as already suggested by Fig. \ref{fig_all_bbh}, with two exceptions.
First, one BBH that is predicted to merge if the simulation is run only for $100 \, \mathrm{Myr}$, is later disrupted by dynamical interactions, and no longer exists at $1500 \, \mathrm{Myr}$. 
Also, the { second} most massive merger (with a final remnant mass $m_{\mathrm{tot}}=99 \, \mathrm{M}_\odot$) needs to dynamically harden for longer than $100 \, \mathrm{Myr}$ to enter the regime in which the orbital decay by GWs becomes effective. As shown in Fig. \ref{fig_bin_form} (lower panel), { 70\%} of the { binaries that give birth to} merging BBHs form at  early stages, when  both components are stars. At later stages, the scarce efficiency of dynamical hardening in low-mass clusters quenches the formation of further BBH mergers. 

The properties of BBH mergers are summarized in Table~\ref{tab_BBHmergers}. 
In low-mass clusters, { almost} all the BBHs are no longer bound to their host cluster when they merge. 
{ In this work, a BBH merger is labelled as bound 
if it merges inside the cluster during the simulation. }

\begin{table}
  \centering
    \begin{tabular}{lrrrr}
    \hline
          & \multicolumn{2}{c}{Low-mass clusters} & \multicolumn{2}{c}{High-mass clusters} \\
          BBH mergers & \multicolumn{1}{l}{100 Myr} & \multicolumn{1}{l}{1500 Myr} & \multicolumn{1}{l}{100 Myr} & \multicolumn{1}{l}{1500 Myr} \\
          \hline
          \hline
    
    All  &  40     &  40     &  115   & 307  \\
    Original &   30    &    30   &    64   & 60  \\
    Exchanged &  10     &    10   &  51  &  247  \\
    Inside YSC  &    1   &   1    &   25    &  174  \\
    IMBHs  &    1   &   1    &  4   &  47  \\
    $m_{\mathrm{tot, max}} \, [\mathrm{M_{\odot}}]$   &    126   &   126    &  249    &  273  \\
    \hline
    \end{tabular}  
  \caption{{ We report the number of all (first row), original (second row) and exchanged (third row) BBH mergers. We also show the number of BBHs that merge inside the cluster during the simulation (fourth row) 
  Finally, we report the number of IMBHs produced by BBH mergers (merger remnants, fifth row), and their maximum BH mass (last row).} 
  }\label{tab_BBHmergers}
\end{table}%

\subsubsection{High-mass clusters}

{  High-mass star clusters host a population of BBH mergers about { eight} times larger than low-mass clusters, although the total initial mass of the  two sets of star clusters is approximately the same.}
This enhancement of BBH mergers in high-mass clusters is particularly evident for the exchanged systems, which, at $1500 \ \mathrm{Myr}$, represent the majority of BBH mergers.

The populations of BBH mergers at 100 and 1500 Myr show notable differences. 
A number of original BBHs that, at 100 Myr, are predicted to merge  are later disrupted (Figure \ref{fig_bbh_merger} and Table~\ref{tab_BBHmergers}). 
Some exchanged BBHs are also disrupted after the first $100 \, \mathrm{Myr}$. However, these disrupted exchanged BBHs are compensated by the  late formation and/or hardening of other exchanged BBHs: we predict { 51  exchanged BBH mergers} at 100 Myr {($\sim 4 \, t_{\mathrm{rlx}}$)}, while at 1500 Myr {($\sim 60 \, t_{\mathrm{rlx}}$)} we find { five times more exchanged BBH mergers}, as shown in Table~\ref{tab_BBHmergers}.
{ Figure~\ref{fig_bin_form} shows that most of the  binaries that result in BBH mergers} 
form { via exchange} when  both components have already { collapsed} to BHs. As opposed to low-mass clusters, merging BBHs can form at very late stages, up to $1000 $ Myr. 

{ A large number of BBHs (174) merge during our simulations,  while they are still inside their parent cluster. Their post-merger remnant is always ejected from the cluster by gravitational recoil (see Sect. \ref{sec_rel_kick}), thus preventing  the possibility of second-generation BH mergers. }

In high-mass clusters, { 47 BBH mergers} give birth to IMBHs, with a remnant mass $m_{\mathrm{tot}}> 100 \, \mathrm{M_{\odot}}$. In the { eight} most massive mergers, the primary BH is itself an IMBH. The most massive merger remnant has a { mass $m_{\mathrm{tot}}=273 \, \mathrm{M_\odot}$.}

\begin{figure}
\begin{center}
\includegraphics[width=\hsize]{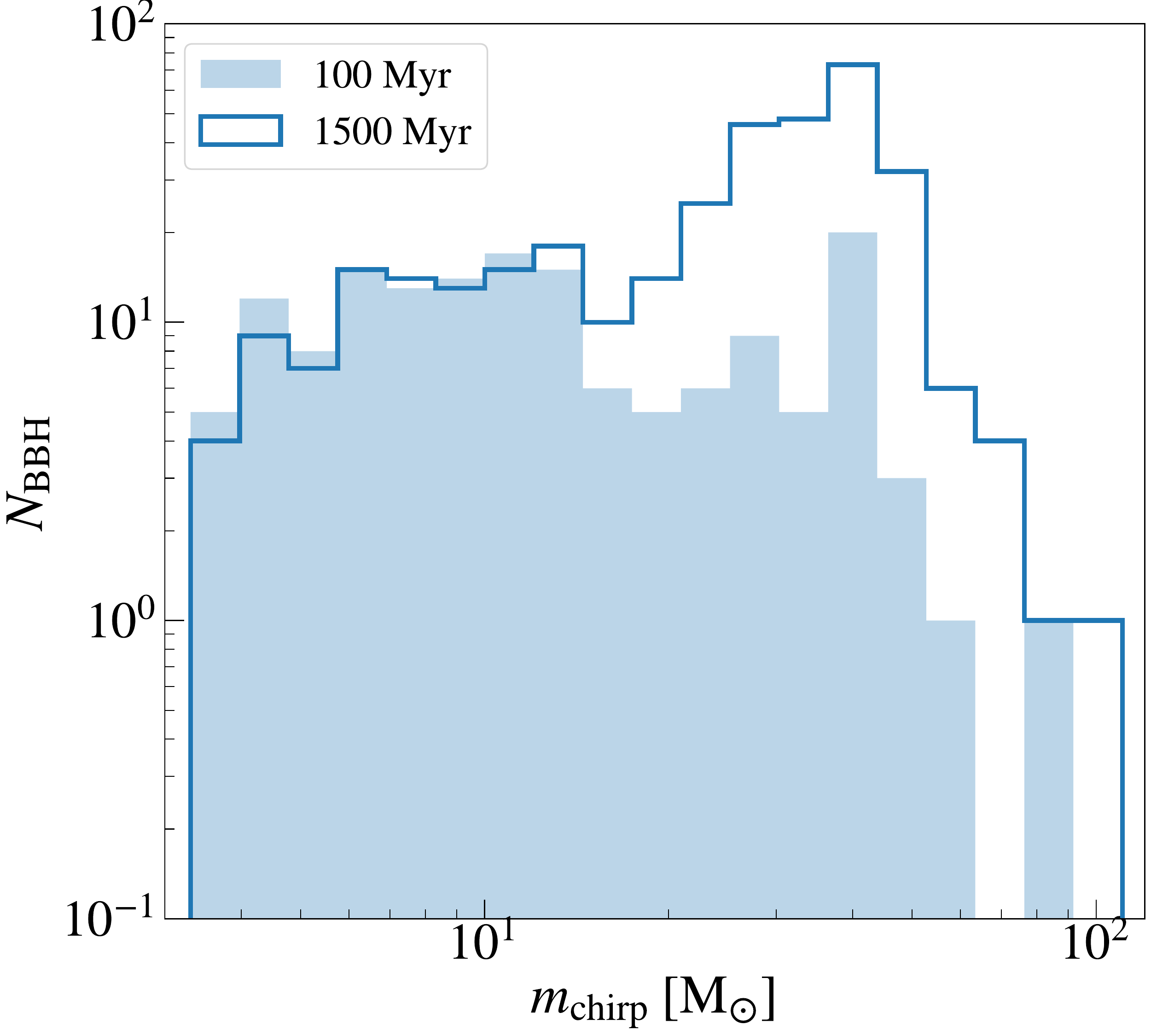}
\caption{{  Chirp mass distribution of merging BBHs at 100 Myr (light-blue filled histogram) and at 1500 Myr (blue line), for all the simulated clusters.}}\label{fig_chirp}
\end{center}
\end{figure}

Figure \ref{fig_chirp} shows the distribution of chirp masses of BBH mergers, for the two considered snapshots (100 and 1500 Myr). The changes in the distribution are mostly due to the long-term dynamical activity within high-mass clusters. The late dynamical activity triggers a large increase of the number of mergers with high chirp mass $m_{\rm chirp}\approx 35-40 \, \mathrm{M_{\odot}}$.

\begin{figure*}
\begin{center}
\includegraphics[width=\textwidth]{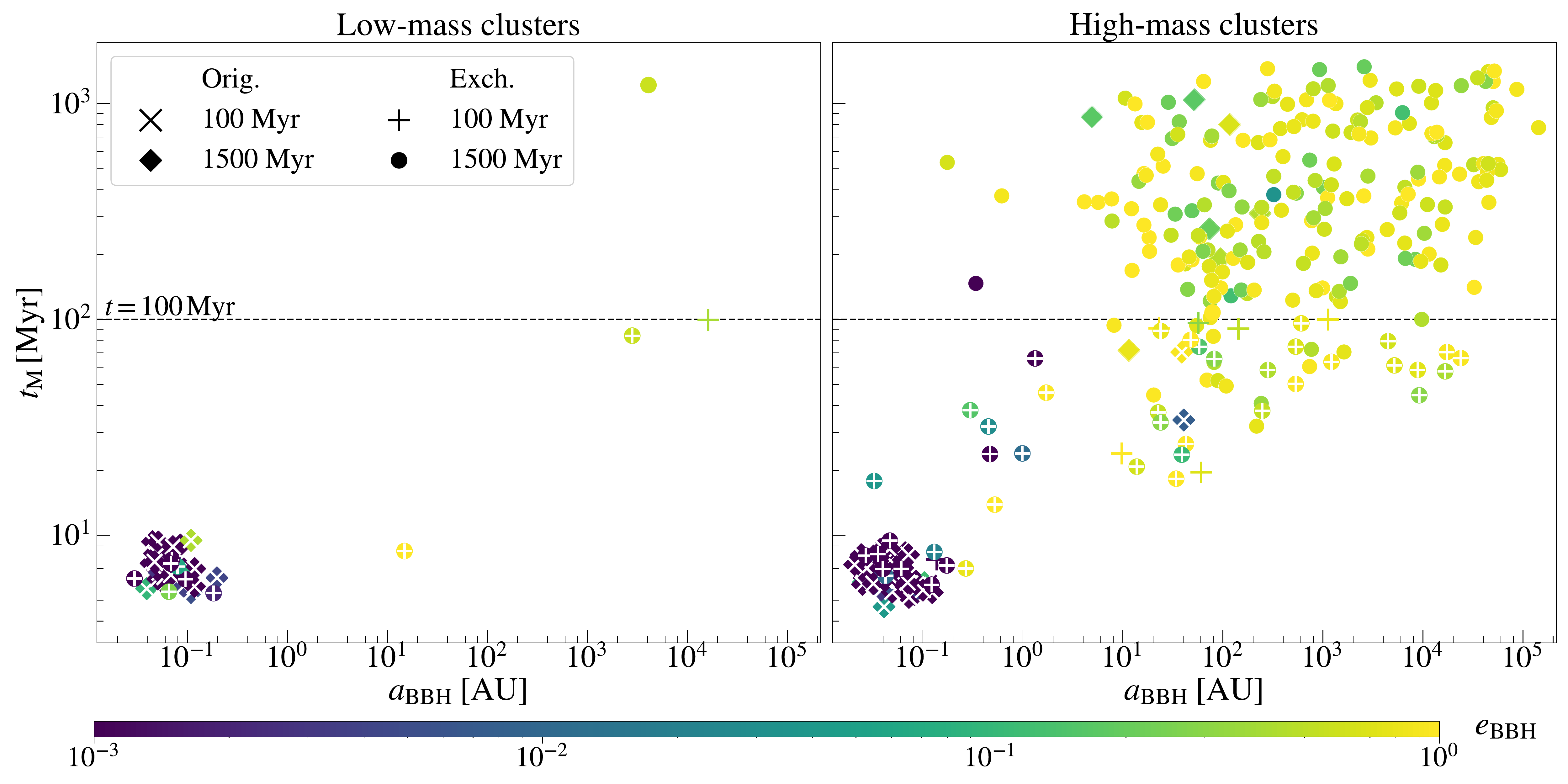}
\caption{
{ Time at which the semi-major axis of the BBH has become sufficiently tight to merge within a Hubble time via GW emission (according to \protect{\citealt{peters1964}}) versus semi-major axis of the BBH when it forms ($a_{\mathrm{BBH}}$), for merging BBHs in low-mass clusters (left) and high-mass clusters (right). The markers are the same as in Figure \ref{fig_bbh_merger}. The colour-map encodes the information on the orbital eccentricity at the BBH formation, $e_{\mathrm{BBH}}$. If a BBH at 1500 Myr is also present at 100 Myr, it is marked with a white cross (original) or plus (exchanged).}}\label{fig_a0_e0}
\end{center}
\end{figure*}


\begin{figure*}
\begin{center}
\includegraphics[width=\textwidth]{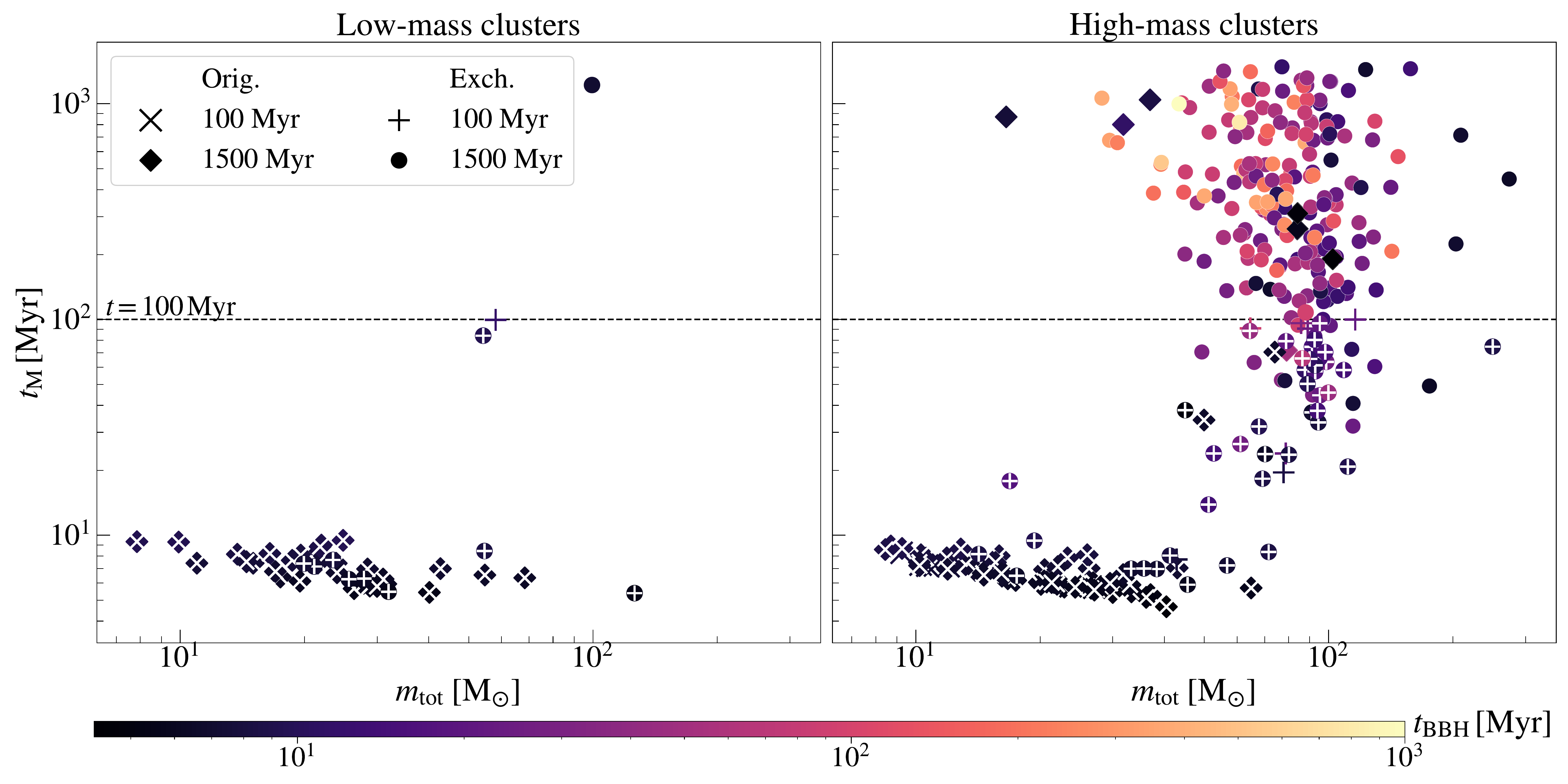}
\caption{
{ Time at which the semi-major axis of the BBH has become sufficiently tight to merge within a Hubble time via GW emission (according to \protect{\citealt{peters1964}}) versus total mass of the BBH merger ($m_{\mathrm{tot}}$), for merging BBHs in low-mass clusters (left) and high-mass clusters (right). The markers are the same as in Figure \ref{fig_bbh_merger}. The colour-map encodes the information on the time at which the BBH forms, $t_{\mathrm{BBH}}$. If a BBH at 1500 Myr is also present at 100 Myr, it is marked with a white cross (original) or plus (exchanged).}}\label{fig_mtot_tM}
\end{center}
\end{figure*}


\subsection{BBH orbital properties at formation}

To estimate for how long a stellar cluster is dynamically active and can affect the formation of BBH mergers, we evaluated $t_{\mathrm{M}}$, defined as the time (since the beginning of the simulation) at which the semi-major axis of the BBH has become sufficiently tight to merge within a Hubble time via GW emission. 
Figure \ref{fig_a0_e0} shows $t_{\mathrm{M}}$ as a function of the  { initial} orbital properties of the BBH, that is its { initial} semi-major axis ($a_{\mathrm{BBH}}$) and orbital eccentricity ($e_{\mathrm{BBH}}$).

In both low-mass and high-mass clusters, the original BBH mergers show typical values of $t_{\mathrm{M}} \lesssim 10 \, \mathrm{Myr}$,  $a_{\mathrm{BBH}} \lesssim 0.1 \, \mathrm{AU}$, and circular orbits. These properties spring from their formation pathway. These BBHs are, in fact, the result of original binaries that hardened as a consequence of a common envelope phase. When the second BH forms, the orbital properties of the BBH already allow it to merge within an Hubble time. For this class of BBH mergers, then, $t_{\mathrm{M}}$ mainly coincides with the time at which the second BH in the binary forms. Also, because the common envelope phase leads to a large mass loss, the resulting BH masses are systematically smaller than the exchanged ones.

As a confirmation of this idea, 
Fig. \ref{fig_mtot_tM} shows $t_{\mathrm{M}}$ as a function of the total mass of the merging BBH, $m_{\mathrm{tot}}$ and the time at which the BBH forms, $t_{\mathrm{BBH}}$. Original BBHs have  $t_{\mathrm{BBH}}\lesssim 10 \,{\mathrm{Myr}}$ and, in most cases, $t_{\mathrm{BBH}}=t_{\mathrm{M}}$. { In high-mass stellar cluster, seven original BBHs show $t_{\mathrm{M}}>t_{\mathrm{BBH}}$, with $t_{\mathrm{M}}$ that can be as high as 1000 Myr. In these cases, dynamical hardening allows the binary system to enter the GW regime after the BBH formation. Because these mergers have not undergone a common envelope phase, they can have masses comparable to the exchanged BBHs (up to $m_{\mathrm{tot}}=102 \, \mathrm{M_{\odot}}$).}
In contrast, $t_{\mathrm{BBH}}$ ranges from 5 Myr to 1100 Myr for exchanged BBHs. In high-mass clusters, { more than 20\% of all BBHs} form after the first 100 Myr. 
Because exchanged BBHs have not undergone mass loss by a common envelope phase, { and because dynamical exchanges favour the formation of massive binaries,} their total masses are systematically higher than those of original BBH mergers, with $m_{\mathrm{tot}} \gtrsim 40 \, \mathrm{M_{\odot}}$.

In low-mass YSCs, only { three} exchanged BBH mergers have 
$a_{\mathrm{BBH}} \gtrsim 1$ AU. In two cases, these mergers correspond to the two most massive BBHs, which formed in dynamically active environments. As a further proof of their dynamical origin, these BBHs are characterized by eccentric orbits.
In high-mass clusters, where dynamical interactions play a major role, the distribution of BBH mergers extends to higher values of $a_{\mathrm{BBH}}$ and $t_{\mathrm{M}}$.
In particular, exchanged binaries, when they form, are generally characterized by large  semi-major axes, up to $1.5 \times10^{4} \, \mathrm{AU}$, and thus take longer times to enter the regime in which GWs efficiently shrink the semi-major axis. In some cases, $t_{\mathrm{M}}$ can be as high as $1400 \, \mathrm{Myr}$, indicating that dynamical hardening can play a role even at the very end of the simulation.

Finally, the dynamical encounters that lead to the formation of BBHs leave a distinctive imprint on their eccentricity. The resulting binary systems are, in fact, characterized by larger eccentricities at formation, with $e_{\mathrm{BBH}}>0.1$\footnote{{ In this discussion, we refer to the eccentricity at the BBH formation. During the in-spiral phase, the BBH mergers will still be circularized as a consequence of GW emission.}}. Exchanged BBHs that have values of $t_{\mathrm{M}} \lesssim 100 \, \mathrm{Myr}$ and high eccentricities can be later disrupted by dynamical interactions, and are no longer present at 1500 Myr. 


\subsection{Formation pathway of BHs in the upper mass gap}

\begin{figure}
\begin{center}
\includegraphics[width=\hsize]{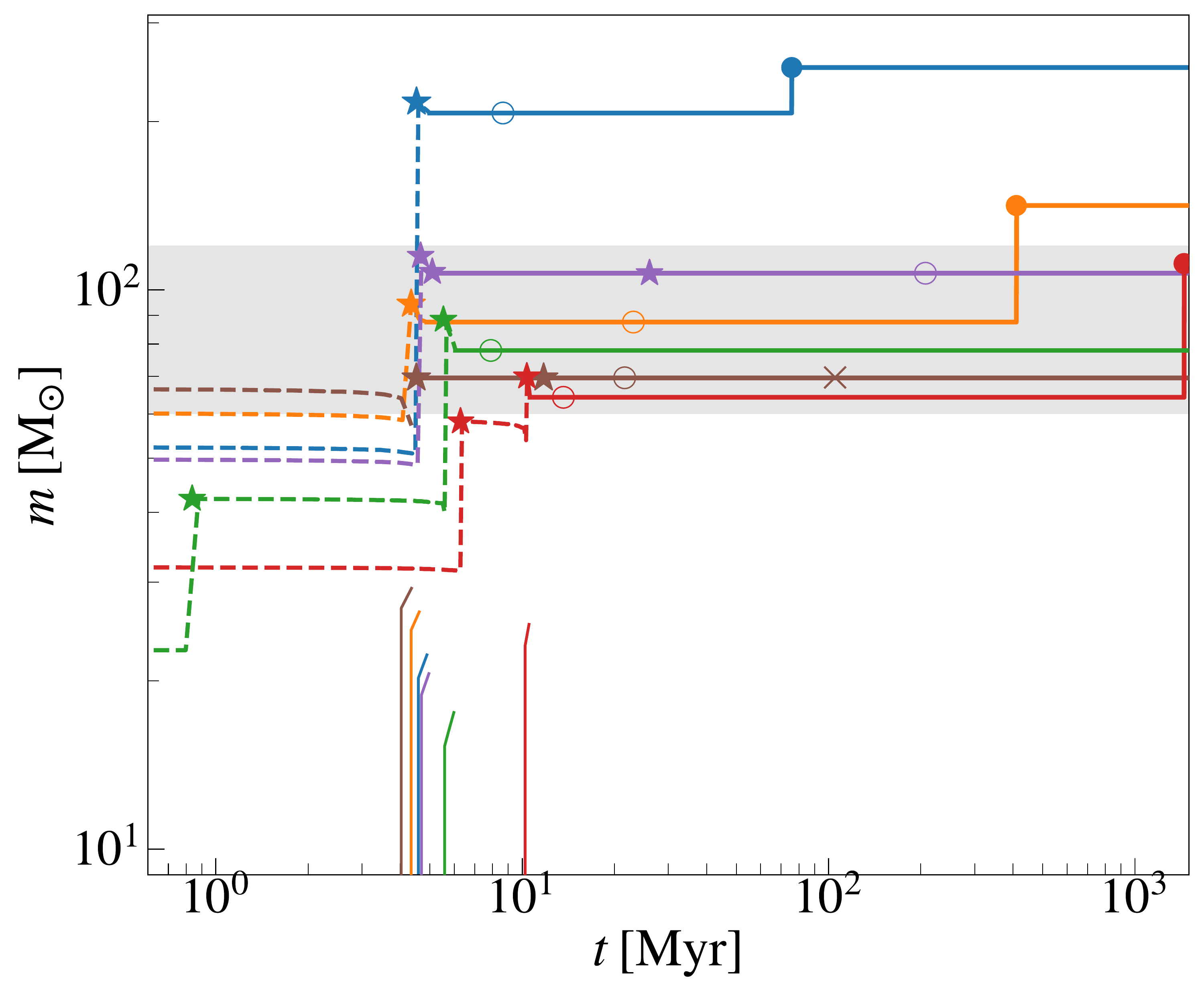}
\caption{{  Evolution of the total mass (thick lines) and  core
mass (thin lines), for the progenitors of BHs
in the PI mass gap and for the most massive primary component of a  BBH merger (blue line). The dashed lines mark the time interval before the star becomes a BH. 
Different markers indicate: the merger between the progenitor star or the BH and another star (stars), the time when the BBH forms (open circles), the merger between the BH and another BH (filled circle), and the time at which the binary is possibly disrupted (crosses). The grey area encloses the PI mass gap, from $\sim 60 \, \mathrm{M_{\odot}}$ to $\sim 120 \, \mathrm{M_{\odot}}$. }}\label{fig_PI}
\end{center}
\end{figure}

{ In high-mass (low-mass) stellar clusters, the primary component of $26$  BBH mergers (1 BBH merger) has mass in the PI  gap. This corresponds to 8\% (2.5\%) of all BBH mergers in high-mass (low-mass) clusters.} 
Figure \ref{fig_PI} shows the evolution of { some} BHs in the PI mass gap  { that become the primary components of BBH mergers}. In all cases, the progenitor star undergoes at least one { collision} with another star. The merger product of such stellar collisions is an exotic star, with an undersized He core with respect to the hydrogen-rich envelope. Such star does not develop PI, because its central properties (temperature and density) do not fall within the PI regime  \citep[e.g.,][]{renzo2020b,costa2021,costa2022,ballone2022}. At the end of its evolution, the stellar product directly collapses into a BH more massive than $60 \, \mathrm{M_{\odot}}$.

{ In all 
our simulations, the binary system that eventually results in a BBH merger with primary mass in the PI gap forms via dynamical exchanges, when both components have already collapsed into BHs. 
 We conservatively assume that mergers between a BH and a star do not affect the mass of the BH, because we expect mass accretion onto the BH to be very inefficient  (\citealt{dicarlo2020,dicarlo2020b}, but see \citealt{rizzuto2021} for a different assumption).} 

\section{Summary} \label{sec_conclusions} 

We have studied the formation of BBHs in young and open star clusters via direct $N$-body simulations, exploiting the codes {\sc nbody6++gpu} \citep{wang2015} and {\sc mobse} \citep{mapelli2017}.  We simulated two different classes of star clusters: low-mass (500--800 M$_\odot$) and { relatively} high-mass (5000--8000 M$_\odot$) systems. We find that the properties and timescales of BBH mergers in the two  star-cluster families are extremely different.

In low-mass clusters, most BBHs form in the first 100 Myr and are the result of the evolution of original binary stars, 
which evolve through common envelope. They do not harden significantly after $\sim{100}$ Myr.
In contrast, the late evolutionary stages ($>1$ Gyr) are crucial for high-mass clusters. Exchanged BBHs (i.e., BBHs that form via dynamical exchanges) are the most common BBH mergers in high-mass clusters (Figures~\ref{fig_all_bbh} and \ref{fig_bbh_merger}). While exchanged BBHs form preferentially in the first $\sim{100}$~Myr, they keep hardening significantly until the end of the simulations (1.5 Gyr, Figure~\ref{fig_bin_form}). This confirms the importance of integrating the evolution of relatively massive clusters ($\gtrsim{}5000$ M$_\odot$) for $>1$ Gyr. 




This difference between the BBH population of low-mass and high-mass star clusters mostly springs from the different two-body relaxation timescale and tidal disruption timescale of the two star cluster families. Our low-mass and high-mass star clusters have { an average} 
two-body relaxation timescale \citep{spitzer1987} of $\sim{10}$ Myr and $\sim{26}$ Myr, respectively.  
This means that mass segregation and other dynamical processes happen earlier in low-mass clusters.  Furthermore, low-mass clusters { dissolve  already at  $\sim{300}$ Myr because of the galactic tidal field, while our high-mass clusters become tidally filling at $\gtrsim{1200}$ Myr} (Figure~\ref{fig_radii}). Hence, the dynamical activity of the low-mass clusters is quenched by tidal evaporation about { four} times earlier than that of high-mass clusters.


In both low-mass and high-mass clusters, the latest BBHs that form (exchanged BBHs) are the most massive ones (primary mass $\gtrsim{}30$ M$_\odot$), because dynamical exchanges favour the pairing of the most massive BHs (Figure~\ref{fig_bbh_merger}). The distribution of the chirp mass of BBH mergers shows two main peaks: 
the { main peak} 
at $\sim{30-40}$ M$_\odot$, and { a secondary peak}  at $\sim{7-15}$ M$_\odot$. { The high-mass peak develops mainly
after 100 Myr (Figure~\ref{fig_chirp}).}

{ These results confirm that we must integrate the evolution of a star cluster for at least 50 two-body relaxation timescales if we want  to probe its BBH population.}


BBH mergers in low-mass clusters are driven mostly by binary evolution via common envelope: they form with short semi-major axis ($\sim{0.1}$~AU) and low orbital eccentricity (Figure~\ref{fig_a0_e0}). In contrast, massive BBHs in high-mass clusters form with larger semi-major axis ($>10$~AU) and higher orbital eccentricity ($0.1-1$).


A non-negligible percentage {($8\%$)} of our simulated BBH mergers in high-mass clusters have primary component's mass in the pair-instability (PI) mass gap. All of them form via stellar collisions, in which a main-sequence star merges with a more evolved star (core He burning). About { 80\%} of these massive BBHs leave a merger remnant in the IMBH range. { In contrast, in low-mass clusters only one dynamical BBH merger produces an IMBH.} 


Furthermore, in the high-mass clusters, we find a few original BBHs with primary mass in the PI mass gap. These are  systems in which one of the two components of the binary star undergoes a collision with a third star and collapses to a BH in the PI mass gap without leading to the ionization of the original binary system.

Overall, our study shows that the formation channels of BBHs in low-mass ($\sim{500-800}$ M$_\odot$) and high-mass star clusters ($\ge{5000}$ M$_\odot$) are extremely different and lead to two completely distinct BBH populations. Low-mass clusters host mainly low-mass BBHs born from binary evolution, while BBHs in high-mass clusters are relatively massive and driven by exchanges. This difference is crucial for the interpretation of GW  sources.



\section*{Acknowledgements}
{ We thank the anonymous referee for the insightful comments, which helped to improve the quality of this manuscript.}
MM, AB and SR knowledge financial support from the EuropeanResearch Council for the ERC Consolidator grant DEMOBLACK, under contract no. 770017.
We thank Nicola Giacobbo and the members of the DEMOBLACK team for useful discussions. We acknowledge that the results of this research have been achieved using the DECI resource Snellius based in the Netherlands at SURFsara, with
support from the PRACE aisbl.
MP acknowledges financial support from the European Union’s Horizon 2020 research and innovation programme under the Marie Sklodowska-Curie grant agreement No. $896248$. ST thanks Mark Gieles and the ICCUB Virgo team for useful comments and discussions.

\section*{Data availability}
The data underlying this article will be shared on reasonable request to the corresponding authors.  The latest public version of {\sc mobse} can be downloaded from \href{https://gitlab.com/micmap/mobse_open}{this repository}.



\bibliographystyle{mnras}
\bibliography{bibliography} 

\bsp	
\label{lastpage}
\end{document}